\documentstyle[aps,12pt,amssymb]{revtex}
\begin{document}



\newcommand{\ethbar}{\bar{\eth}}
\newcommand{\scrip}{\cal I^+} \newcommand{\scripm}{\cal I^-}
\newcommand{\scri}{\cal I}
\newcommand{\ihat}{\hat{\imath}} \newcommand{\jhat}{\hat{\jmath}}
\newcommand{\omegabar}{\overline{\omega}}
\newcommand{\deltabar}{\overline{\delta}}
\newcommand{\sigmabar}{\overline{\sigma}}
\newcommand{\Lambdabar}{\bar{\Lambda}}
\newcommand{\Deltabar}{\bar{\Delta}}
\newcommand{\zetabar}{\bar{\zeta}}
\newcommand{\etabar}{\overline{\eta}}
\newcommand{\Wbar}{\overline{W}}
\newcommand{\Tbar}{\overline{T}}
\newcommand{\Hbar}{\overline{H}}
\newcommand{\Gbar}{\overline{G}}
\newcommand{\Ibar}{\overline{I}}
\newcommand{\lhbar}{\overline{h}}
\newcommand{\Abar}{\bar{A}}
\newcommand{\mbar}{\overline{m}}
\newcommand{\Mbar}{\overline{M}}
\newcommand{\Sbar}{\overline{S}}
\newcommand{\Hspace}{${\cal H}$-space }
\newcommand{\ezero}{e\!\!\!^{^{\scriptscriptstyle{(0)}}}}
\newcommand{\eone}{e\!\!\!^{^{\scriptscriptstyle{(1)}}}}
\newcommand{\etwo}{e\!\!\!^{^{\scriptscriptstyle{(2)}}}}
\newcommand{\Hone}{H\!\!\!\!^{^{^{\scriptscriptstyle{(1)}}}}}
\newcommand{\Htwo}{H\!\!\!\!^{^{^{\scriptscriptstyle{(2)}}}}}
\newcommand{\Hthree}{H\!\!\!\!^{^{^{\scriptscriptstyle{(3)}}}}}
\newcommand{\htwo}{h\!\!\!\!^{^{^{\scriptscriptstyle{(2)}}}}}
\newcommand{\hthree}{h\!\!\!\!^{^{^{\scriptscriptstyle{(3)}}}}}
\newcommand{\Lambdaone}{\Lambda\!\!\!\!^{^{^{\scriptscriptstyle{(1)}}}}}
\newcommand{\Lambdatwo}{\Lambda\!\!\!\!^{^{^{\scriptscriptstyle{(2)}}}}}
\newcommand{\Lambdathree}{\Lambda\!\!\!\!^{^{^{\scriptscriptstyle{(3)}}}}}
\newcommand{\Lambarone}{\Lambdabar\!\!\!\!^{^{^{^{\scriptscriptstyle{(1)}}}}}}
\newcommand{\Lambartwo}{\Lambdabar\!\!\!\!^{^{^{^{\scriptscriptstyle{(2)}}}}}}
\newcommand{\Zzero}{Z\!\!\!\!^{^{^{\scriptscriptstyle{(0)}}}}}
\newcommand{\Zone}{Z\!\!\!\!^{^{^{\scriptscriptstyle{(1)}}}}}
\newcommand{\Ztwo}{Z\!\!\!\!^{^{^{\scriptscriptstyle{(2)}}}}}
\newcommand{\lone}{\Lambda_1} \newcommand{\lonebar}{\Lambdabar_1}
\newcommand{\etatw}{\tilde \eta} \newcommand{\zetatw}{\tilde \zeta}
\newcommand{\bondidotbar}{{\dot{\overline \sigma}}_{\scriptscriptstyle B}}
\newcommand{\bondidot}{{\dot{\sigma}}_{\scriptscriptstyle B}}
\newcommand{\bondibar}{{\overline{\sigma}}_{\scriptscriptstyle B}}
\newcommand{\bondi}{{\sigma}_{\scriptscriptstyle B}}
\newcommand{\be}{\begin{equation}}
\newcommand{\ee}{\end{equation}}
\newcommand{\bea}{\begin{eqnarray}}
\newcommand{\eea}{\end{eqnarray}}

\title{The Vacuum Einstein Equations via Holonomy around Closed Loops
on Characteristic Surfaces.}

\vspace{2cm}

\author{Savitri V. Iyer, \\
{\small Department of Physics and Astronomy, State University of New York,
Geneseo, NY 14544}\\
\vspace{0.5cm}
Carlos N. Kozameh,\\
{\small FaMAF, University of Cordoba, 5000 Cordoba, Argentina}\\
\vspace{0.5cm}
Ezra T. Newman,\\
{\small Department of Physics, University of Pittsburgh, Pittsburgh, PA 15250}}

\maketitle

\vspace{1cm}

\begin{abstract}

{\bf Abstract}

{\small We reformulate the standard local equations of general relativity for
asymptotically flat spacetimes in terms of two non-local quantities,
the holonomy $H$ around certain closed null loops on characteristic
surfaces and the light cone cut function $Z$, which describes the
intersection of the future null cones from arbitrary spacetime points,
with future null infinity.}

{\small We obtain a set of differential equations for $H$ and $Z$ equivalent to
the vacuum Einstein equations.  By finding an algebraic relation
between $H$ and $Z$ this set of equations is reduced to just two
coupled equations: an integro-differential equation for $Z$ which
yields the conformal structure of the underlying spacetime and a linear
differential equation for the ``vacuum" conformal factor.  These
equations, which apply to all vacuum asymptotically flat spacetimes,
are however lengthy and complicated and we do not yet know of any
solution generating technique.  They nevertheless are amenable to an
attractive perturbative scheme which has Minkowski space as a zeroth
order solution.}

\end{abstract}
\newpage
\section{Introduction.}

The vacuum Maxwell fields on Minkowski space can be expressed as
Kirchhoff integrals taken over initial data, i.e., data given on a
Cauchy surface.  If we replace the spacelike surface by a
characteristic, or null, surface, we get the analog to the Kirchhoff
integral, the D'Adhemar integral \cite{penrose}.  Maxwell and
Yang-Mills fields on both flat and asymptotically flat spacetimes have
been given such D'Adhemar-like reformulations with considerable success
\cite{Max,YM,Curved}.  The present work is an attempt to extend this
technique to general relativity.  We derive differential equations
whose solutions would allow us to express the gravitational field at
any given spacetime point as a series of D'Adhemar-like integral over free data
given on a characteristic surface.

First consider the well understood analog to our problem, determining
the Maxwell field at some interior point on Minkowski space.  Free data
is given on the null surface $\scri$, which is the three dimensional
boundary of the spacetime manifold $\cal M$.  The future light cone
emanating from a given interior point $x^a$ would intersect the future
half $\scrip$ of $\scri$, the intersection being a 2-surface.  We refer
to this 2-surface as the ``light cone cut" of $\scrip$associated with
$x^a$.  The D'Adhemar formulation of the Maxwell equations involves
integrals over the free data given on the light cone cut.  It is
crucial that the light cone structure of the underlying manifold be
known, in order to perform this integration.  Although for the
Minkowski metric this is readily available, it takes considerable
effort to obtain this information for an {\it arbitrary} background.  A
generalization of the D'Adhemar formulation of the Maxwell field in
Minkowski space to the case when the background is curved is available
\cite{Curved}, although the resulting equations become complicated.  A
further generalization of this technique to Yang-Mills fields on both
flat and asymptotically flat spacetimes are also available
\cite{YMcurved}.  In these (Yang-Mills) cases, in addition to the above
mentioned difficulty in the determination of the light cones, the final
equations are further complicated by the introduction of
non-linearity.  The D'Adhemar integrals (for the Yang-Mills case and the
Maxwell case on a curved background) shows explicitly that propagation
becomes non-Huygens--i.e., propagation is not confined to the light
cone.

A D'Adhemar-like formulation for general relativity might seem
inaccessible since the gravitational field is not separable from the
background and its characteristic surfaces: it requires {\it
simultaneously} solving for the field and the null surfaces.  It is
nevertheless possible to give such a formulation by dividing the
equations into two parts: one resembling the D'Adhemar integrals for a
(self-dual or anti-self-dual) Yang-Mills field, and the other yielding
the null surfaces that are needed to give meaning to the integrals in
the first part. The two parts are strongly coupled.

The first set of equations is equivalent to the D'Adhemar version of
the self-dual and the anti-self- dual Yang-Mills equations for the
self-dual and anti-self-dual parts of an $O(3,1)$ connection,
respectively, on a spacetime manifold $\cal M$ with an arbitrary
asymptotically flat metric.  The basic idea is to treat a special case
of the $O(3,1)$ Yang-Mills connection so that it agrees with that of
the background gravitational connection.  This is done by imposing very
restrictive conditions on the Yang-Mills data.

The second set of equations determines the light cones of the manifold
$\cal M$.  For a {\it given} metric, one could have obtained this
information by integrating the null geodesic equations (see for example
\cite{schwld}).  However, in our case where the metric is as yet
undetermined, we derive differential equations for the characteristic
surfaces that are coupled to the $O(3,1)$ Yang-Mills field equations.

The above two parts, which look hopelessly intertwined at first,
nevertheless give results that are tractable and comprehensible.  The
exact equations, while complicated, are in principle solvable.  An
attractive feature of this approach is that the equations are amenable
to an approximation scheme wherein the $n$-th order correction to the
field is expressible as D'Adhemar integrals in terms of lower order.

As the material of this paper is not at all close to mainstream ideas,
we believe that it might be appropriate first to give a broad
perspective before going into the details.

There are two different sets of ideas that we try to weld together.
The first is based on the use of characteristic surfaces on arbitrary
asymptotically flat Lorentzian spacetimes.  We make essential use of
the future light cones $N_x$ from arbitrary spacetime points $x^a$ and
their intersections, $C_x$, with future null infinity, $\scrip$,
(referred to as the light cone cuts)--as well as the past light cones
from points on $\scrip$. These characteristic surfaces are used to give
the D'Adhemar-like reformulation of the Yang-Mills equations.  Using
Bondi coordinates $(u, \zeta, \zetabar)$ on $\scrip$, $C_x$ is
described by a function referred to as the light cone cut function on
$\scrip$ of the form $u=Z(x^a, \zeta, \zetabar)$, parametrized by the
interior points $x^a$ (see Figures 1 and 2).  The cut function plays a
dual role: for a fixed interior point $x^a$ it describes the ``cut" of
$\scrip$, and for a fixed point $(u, \zeta, \zetabar)$ on $\scrip$, it
describes the past light cone of that point as $x^a$ is varied. From the
latter point of view the cut function becomes a three-parameter family
of characteristic surfaces. The conformal structure of spacetime can be
recovered
from knowledge of the cut function, i.e., the conformal metric can
be explicitly written in terms of derivatives of $Z$. More specifically, the
conformal metric can be completely expressed in terms of $\Lambda$, defined
from $Z$
by $\Lambda \equiv \eth {}^2 Z$. The quantity  $\Lambda$, becomes one of our
basic
variables. These ideas are developed in Section II.

The second set of ideas of concern to us (see Section III) is based on
the fact that, in some sense, the $O(3,1)$ Yang-Mills equations
contain, as a special class of solutions, the vacuum Einstein
equations.  The full Yang-Mills curvature, $F$, can be broken into four
pieces; self-dual and anti-self-dual on the internal indices as well as
self-dual and anti-self-dual on the spacetime indices, so that
\be
F={}^+F^+ + {}^+F^- + {}^-F^+ + {}^-F^-,
\ee
where the ${+,-}$ symbols to the left refer to spacetime duality and those to
the right refer to internal
duality.  The Einstein equations are almost encoded\cite{Lamberti} into the
algebraic
statement that ${}^+F^- = {}^-F^+ = 0$ needing only an appropriate
restriction of the data.  The basic idea is to think of solving these
equations (for the self-dual and the anti-self-dual fields ${}^+F^+$
and ${}^-F^-$ and their respective connections) on a given
asymptotically flat vacuum background metric, then introduce a soldering
form and restrict the data so that it coincides with the background
connection data.  An important technical device is to give these
equations a D'Adhemar-like formulation which, as was mentioned earlier,
can be done for Yang-Mills fields.  The tool for this is the use of the
holonomy operator H (around special null paths) as the primitive variable
for the Yang-Mills fields.  The Bianchi identities, equivalent (because
of the self- and anti-self-duality) to the Yang-Mills field equations become
one set of our
final equations. They are symbolically written as equations for H as

\be {\cal B}_\beta (H, A) = 0, \label{symbbi}
\ee
with $A$ being the restricted characteristic data.

When the soldering form (a tetrad field) is introduced, we can obtain a set of
relationships between the spacetime variable $ \Lambda \equiv \eth {}^2 Z$, and
$H$, that
we refer to as the $\Lambda(H)$-relations given symbolically by

\be
\Lambda = \Lambda(H).
\ee
Our third set of equations, referred to as the ``field equations'', are simply
the algebraic
equations

\be
{}^+F^- = {}^-F^+ = 0,
\ee
expressed as functions of the
components of the holonomy operator, symbolically written as
\be
{\cal
D} (H) = 0.\label{symbh}
\ee

Our final task is two-fold: To

\begin{enumerate}
\item eliminate the
holonomy variables from these three relationships and be left with
equations {\it only} for the characteristic surfaces, i.e., equations
for $Z$ (Section III.E.)  These equations constitute the conformal
Einstein equations and yield an equation that we refer to as the light
cone cut equation (LCCE) which involves only $\Lambda$ and the free
data, and
\item find an equation, derivable from equation (\ref{symbh}), for the
conformal factor that converts the conformal metric to a vacuum
metric.
\end{enumerate}

In Section IV we discuss the linearization of
the theory and a perturbation scheme.


\section{Light Cone Cuts.}

Let us consider a real, four-dimensional, asymptotically flat
spacetime, $\cal M$, with a conformal metric $g_{ab}$ and future null
boundary $\scrip$. We use Bondi coordinates $(u, \zeta, \zetabar)$ to
coordinatize $\scrip$.  A coordinate grid of the ($u=$const.) and the
($\zeta, \zetabar =$ const.) surfaces on $\scrip$ is shown in Figure 1.

Consider the future light cone $N_x$ emanating from an internal point
$x^a$.  A relation among the coordinates $(u, \zeta, \zetabar)$ such as
$u=f(\zeta, \zetabar)$ describes locally a 2-surface which we refer to
as a `cut'.  The intersection of $N_x$ with $\scrip$ is a special cut,
the `light cone cut' $C_x$ (Figure 2), given by

\be
u=Z(x^a, \zeta, \zetabar). \label{lccf2}
\ee

We refer to $Z(x^a,
\zeta, \zetabar)$ as the light cone cut function.  The function $Z(x^a,
\zeta, \zetabar)$ is our basic variable.  The conformal metric can be
given as an explicit function of $Z$\cite{cnk}.

One could also interpret equation (\ref{lccf2}) as a description of the
{\it past} light cone of the point $(u, \zeta, \zetabar)$ on $\scrip$
(also shown in Figure 2).  That is, keeping $u$, $\zeta$ and
$\zetabar$ fixed, if we vary $x^a$, this equation is the locus of all
spacetime points $x^a$ that are null connected to the point $(u, \zeta,
\zetabar)$, which by definition is the past light cone of $(u, \zeta,
\zetabar)$ on $\scrip$.  From this important observation, we see that
for fixed values of $(u, \zeta, \zetabar)$, $\nabla_a Z$ is a null
covector, i.e.,
\be
g^{ab}{Z,}_a(x,\zeta,\zetabar){Z,}_b(x,\zeta,\zetabar) = 0.\label{nullvect}
\ee

Thus it follows that $Z_{,a}(x, \zeta, \zetabar)$ at the point
$x^a$ sweeps out the null cone at that point as we vary $\zeta$ and
$\zetabar$.  Though it has been discussed elsewhere \cite{cnk} and we
will not go into the details, it is from equation (\ref{nullvect}), by
taking several $\zeta$ and $\zetabar$ derivatives, the entire conformal
metric can be reconstructed completely in terms of $Z$ or more specifically
$\Lambda \equiv \eth {}^2 Z$.

The sphere of null directions at the point $x^a$ is coordinatized by
$\zeta$ and $\zetabar$ .  We use these coordinates for $S^2$ instead of
the usual $(\theta, \phi)$ because covariant differentiation on the
sphere, which appears in many of the equations, takes on a particularly
simple form in terms of $(\zeta, \zetabar)$.  We introduce the
operators $\eth$ and $\ethbar$ \cite{eth}

\be
\eth \alpha_s =
{(1+\zeta \zetabar)}^{1-s} {\partial \over {\partial \zeta}} \left[
{(1+\zeta \zetabar)}^{s} \alpha_s \right]
\ee
and
\be
\ethbar \alpha_s
= {(1+\zeta \zetabar)}^{1+s} {\partial \over {\partial \zetabar}}
\left[ {(1+\zeta \zetabar)}^{-s} \alpha_s \right],
\ee
which operate on the quantity $\alpha_s$ defined on the sphere, where $s$, the
spin
weight, is assigned according to how $\alpha$ transforms under a
specific transformation.
(See \cite{ethgroup} for more about this transformation and properties
of $\eth$ and $\ethbar$.)  It helps to
think of $\eth$ and $\ethbar$ loosely as differentiation with respect
to $\zeta$ and $\zetabar$ respectively.

Assuming that we know the light cone cut function, and with the above
definitions of $\eth$ and $\ethbar$, the following set of quantities
are well-defined:
\bea
u &=& Z(x^{a},\zeta,\zetabar), \label{you} \\
\omega &=& \eth Z(x^{a},\zeta,\zetabar), \label{omega}\\
\omegabar &=& \ethbar Z(x^{a},\zeta,\zetabar), \label{omegab} \\
R &=& \eth \ethbar
Z(x^{a},\zeta,\zetabar), \label{arr}
\eea
with $(u, R, \omega, \omegabar)$ having, respectively, spin-weights $(0, 0, 1,
-1)$.  This
set defines a {\it sphere's worth} of coordinate transformations on the
spacetime parametrized by $(\zeta, \zetabar)$. Let
\be
(\theta^0, \theta^1, \theta^{+}, \theta^{-}) \equiv (u, R, \omega, \omegabar).
\ee
With this notation, the $(\zeta, \zetabar)$-dependent coordinate
transformation can be written as
\be
\theta^i = \theta^i (x^a, \zeta,
\zetabar), \label{coordtrans}
\ee
where the indices ${i}, {j}$ will
take on the values $\{ 0, 1, +, - \}$.

We now construct

\be
\eth {}^2 Z \equiv \lambda (x^a, \zeta,\zetabar).
\ee

The quantity $\Lambda$ and its conjugate can be
expressed as functions of the $\theta^i$ by inverting the
transformation (\ref{coordtrans}) and eliminating the $x^a$, to obtain
\be
\Lambda (\theta^i, \zeta, \zetabar) = \eth {}^2 Z. \label{Lambdadefn}
\ee

Later we show that the (conformal) Einstein
equations can be encoded into the choice of $\Lambda$.

With the $\theta^{i}$ coordinate system, we have a one-form basis:
\be
d\theta^{i} = {\theta^{i}}_a d x^a, \;\; {\rm \ with \ } \;\;
{\theta^{i}}_{,a} = {\theta^{i}}_a
\ee
and the dual basis
\be
{\partial \over {\partial \theta^{i}}} = {\theta^a}_{i} {\partial \over
{\partial
x^a}},
\ee
with the relations
\bea
{\theta^{i}}_a {\theta^a}_{j} &=& {\delta^{i}}_{j}, \\
{\theta^{i}}_a {\theta^b}_{i} &=& {\delta_a}^b.
\eea
Any vector $V_a$ can be written as $V_a = V_i {\theta^i}_{a}$ and
in particular we have
\be
\Lambda_{,a} = \Lambda_{,i} {\theta^i}_{a} = \Lambda_{,0} Z_a +\; \; {\rm
etc....}
\ee
The $\Lambda_{,i}$ play a basic role in what follows.

Note also that $\eth {\theta^i}_{a}$ can also be expressed as a linear
combination of the ${\theta^i}_{a}$, i.e., as
\be
\eth {\theta^{i}}_a = {T^{i}}{}_{j} {\theta^{j}}_a \label{etheta}
\ee
{}From ${\theta^i}_{a} \equiv \{ Z_a, \eth \ethbar Z_a, \eth Z_a, \ethbar Z_a
\}$, we obtain, e.g., that

\bea
\eth {\theta^0}_{a} &=& {\theta^+}_{a}, \\
\eth {\theta^+}_{a} &=& \Lambda_{,a} = \Lambda_{,i} {\theta^i}_{a},\\
\eth {\theta^-}_{a} &=& {\theta^1}_{a}, \\
\eth {\theta^1}_{a} &=& {T^1}{}_{i} {\theta^{i}}_a
\eea
with
\bea
{T^1}_{i} &=& {1 \over{q}} \biggl[ \lone(\eth \Lambdabar_i + \Lambdabar_0
{\delta^{+}}_{i} +
\Lambdabar_-{\delta^{1}}_{i} + \Lambdabar_+ \Lambda_i -2
{\delta^{-}}_{i}) \nonumber \\
&& \hspace{1in} + \ethbar \Lambda_i +
\Lambda_0 {\delta^{-}}_{i} + \Lambda_+ {\delta^{1}}_{i} + \Lambda_-
\Lambdabar_i -2 {\delta^{+}}_{i} \biggr], \nonumber
\eea
and
\be
q=(1-\lone \lonebar). \nonumber
\ee

Likewise,
\be
\ethbar {\theta^{i}}_a = {\Tbar^{i}}{}_{j} {\theta^{j}}_a \label{edbartheta}
\ee
with ${\Tbar^{i}}{}_{j}$ obtained from the complex conjugate coefficients of
${T^{i}}{}_{j}$.
The explicit form of these matrices with row and column indices $\{0,1,+,- \}$
are:

\be
{T^{i}}_{j} = \left(
\begin{array}{cccc} 0 & 0 & 1 & 0 \\
{T^1}_0 & {T^1}_1 & {T^1}_{+} & {T^1}_{-} \\
\Lambda_0 & \lone & \Lambda_{+} & \Lambda_{-} \\
0 & 1 & 0 & 0
\end{array}
\right),\label{Tmatrix}
\ee
and
\be
{{\Tbar}^{i}}{}_{j} = \left(
\begin{array}{cccc} 0 & 0 & 0 & 1 \\
{{\Tbar}^1}{}_0 & {{\Tbar}^1}{}_1 & {{\Tbar}^1}{}_{+} & {{\Tbar}^1}{}_{-} \\
0 & 1 & 0 & 0 \\
\Lambdabar_0 & \lonebar & \Lambdabar_{+} & \Lambdabar_{-}
\end{array}
\right),
\ee


\section{Holonomy and Einstein Equations.}

\subsection{$O(3,1)$ Yang-Mills Equations.}

We begin with a real four-dimensional Lorentzian  manifold
$\cal M$ where we {\it assume} that the light cone structure of the
previous Section is known and study an $O(3,1)$ Yang-Mills field. The
idea is to (roughly) think of the associated vector bundle as being the
tangent bundle of the spacetime, though of course a soldering form is
needed to make this precise.

The connection one-form, which is antisymmetric in the Lorentzian
indices, ${\ihat}, {\jhat}$,..., can be decomposed into its self-dual
and anti-self-dual parts,
\be
\gamma_{\scriptscriptstyle a} {}^{\ihat}
{}_{\jhat} = \gamma_{\scriptscriptstyle a}^{+} {}^{\ihat} {}_{\jhat} +
\gamma_{\scriptscriptstyle a}^{-} {}^{\ihat} {}_{\jhat},
\ee
where self-dual and anti-self-dual are defined by
\be
\gamma_{\scriptscriptstyle a}^{\pm} {}^{\ihat} {}_{\jhat} = {1
\over{2}} \left( \gamma_{\scriptscriptstyle a} {}^{\ihat} {}_{\jhat}
\mp i \gamma_{\scriptscriptstyle a}^{\ast} {}^{\ihat} {}_{\jhat}
\right),
\ee
and duality by
\be
\gamma_{\scriptscriptstyle a}^{\ast}
{}^{{\ihat} {\jhat}} = {1 \over{2}} \epsilon^{\hat {\scriptscriptstyle
{\imath}} {\jhat}} {}_{\hat {\scriptscriptstyle k}{\hat \ell}} {\rm \ }
\gamma_{\scriptscriptstyle a}^{\hat {\scriptscriptstyle k} {\hat
\ell}},
\ee
where $\epsilon^{{\ihat} {\jhat}} {}_{{\hat k}{\hat \ell}}$
is the alternating symbol with $\epsilon_{0123} = -1$.  The curvature
tensor can be similarly decomposed as,
\be
F_{ab}{}^{\ihat}{}_{\jhat} =
F_{ab}^{+} {}^{\ihat} {}_{\jhat} + F_{ab}^{-} {}^{\ihat} {}_{\jhat},
\ee
where the self-dual and anti-self-dual parts of $F$ are constructed
from the self-dual and anti-self-dual connections respectively, i.e.,
\be
F_{ab}^{\pm} {}^{\ihat}{}_{\jhat} = \nabla_{\scriptscriptstyle [a}
\gamma_{\scriptscriptstyle b]}^{\pm} {}^{\ihat} {}_{\jhat} +
[\gamma_{\scriptscriptstyle a}^{\pm}, \gamma_{\scriptscriptstyle
b}^{\pm}   {{]}^{\ihat}}_{\jhat}.
\ee

With the above decompositions,
the Bianchi identities and the field equations become
\be
\nabla_{[c} F^{\pm}_{ab]} {}^{\ihat} {}_{\jhat} + {[ \gamma_{[c}^{\pm} ,
F_{ab]}^{\pm} {]}^{\ihat}}_{\jhat} = 0\label{o31bi}
\ee
and
\be
\nabla^a F_{ab}^{\pm} {}^{\ihat} {}_{\jhat} + [\gamma^{\pm} {}^a ,
F_{ab}^{\pm} {]}^{\ihat} {}_{\jhat} = 0, \label{o31fld}
\ee
respectively. One is thus dealing with two independent  Yang-Mills
connections and fields.

It is  possible to further decompose each of the two curvature tensors,
now  on the spacetime indices, into its spacetime self-dual and
anti-self-dual parts, where we have used the existence of the
Lorentzian metric.  We will refer to spacetime duals as left duals and
internal duals as right duals.  The full curvature then has four
terms:

\begin{enumerate}

\item the left and right self-dual part,  $^{+}\!F_{ab}^+$;

\item the left anti-self-dual and right self-dual part,
$^{-}\!F_{ab}^+$;

\item the left self-dual and right anti-self-dual part,
$^{+}\!F_{ab}^-$;

\item the left anti-self-dual and right anti-self-dual part,
$^{-}\!F_{ab}^-$.
\end {enumerate}

Parts (1) and (2) are coupled as
are parts (3) and (4), in the sense that they depend respectively on
the $\gamma^+$ and $\gamma^-$.

We next assume that:
\be
^{-}\!F_{ab}^+ =^{+}\!F_{ab}^- = 0. \label{assume}
\ee

{}From this the field equations (\ref{o31fld}) are automatically
satisfied via the Bianchi
identities (\ref{o31bi}).  Thus we have two independent Yang-Mills
fields: a (left) self-dual field, $^{+}\!F_{ab}^+$ and a (left)
anti-self-dual field, $^{-}\!F_{ab}^-$ satisfying the Bianchi
identities (\ref{o31bi}) and the above condition (\ref{assume}).  These
two equations are rewritten in terms of a new variable, viz., the
holonomy operator, in the next section.

Equations (\ref{o31bi}) and (\ref{assume}) {\it with a special choice
of data} (on $\scrip$) are equivalent to the Einstein equations.


\subsection{Holonomy and the Bianchi Identities.}

The variable that we have been (loosely) referring to as the holonomy
operator and denoting by $H$, is more accurately the difference
between the holonomy operator associated with an infinitesimal path and
the identity.  It should really be called the infinitesimal holonomy
operator or the differential holonomy operator, though we will continue
with the original name.  There are two distinctly different sets of
paths $\Delta_x (\zeta, \zetabar)$ and $\Deltabar_x (\zeta, \zetabar)$
(with their own holonomies $H$ and $\Hbar$) defined in the following
manner:  Consider an interior point $x^a$ of $\cal M$.  For $\Delta_x
(\zeta, \zetabar)$ we choose two null rays on the cone $N_x$ that are
infinitesimally separated, namely,  $\ell_x(\zeta, \zetabar)$ and
$\ell_x(\zeta + d \zeta, \zetabar)$, extending from $x^a$ to $\scrip$
(see Figure 3).  We then form a closed loop by connecting the
end points of these two rays on $\scrip$.  At $\scrip$ the two-form constructed
from  this connecting vector and the tangent vector to the geodesics
 is self-dual.  (In Minkowski space the entire path lies in a self dual blade;
it is the loss of this property in  curved spacetimes that is the source of
non-Huygens
propagation for linear rest-mass zero fields). In a similar manner, one can
choose
the anti-self-dual triangle $\Deltabar_x (\zeta, \zetabar)$, which has
$\ell_x(\zeta, \zetabar)$ and $\ell_x(\zeta + d \zeta, \zetabar)$ for its
sides.

For our $O(3,1)$ connection, the vectors that are being propagated
around the closed loops ($\Delta_x$ and $\Deltabar_x$) are thought of (using a
soldering
form ${\lambda^a}_i (x)$ introduced below)
as being in the tangent (or cotangent) bundle.  The effect of applying
the operators $H$ or $\Hbar$ to an arbitrary vector $V^{\mu}$ at $\scrip$
is, respectively,
\be
{V'}^{\mu} = V^{\nu} \left( {\delta^{\mu}}_{\nu}
+ {{{H^{\mu}}_{\nu} (x^a, \zeta, \zetabar) d \zeta} \over {2 P}}
\right)
\ee
and
\be
{V'}^{\mu} = V^{\nu} \left( {\delta^{\mu}}_{\nu} +
{{{\Hbar^{\mu}}_{\nu} (x^a, \zeta, \zetabar) d \zetabar} \over {2 P}}
\right),
\ee
where the $P =( 1+\zeta \zetabar)$ is for notational
purposes.  $H$ takes us from a point on $\scrip$ along $\ell_x (\zeta,
\zetabar)$ down to $x^a$ then back to $\scrip$ along $\ell_x (\zeta +
d \zeta, \zetabar)$ and finally back to the starting point along a
connecting vector on $\scrip$ (see Figure 3).  For parallel
transport around this loop one obtains the following:
\be
{H^{\mu}}_{\nu} (x^a, \zeta, \zetabar) = {A^{\mu}}_{\nu} +
{{(G^{-1})}_{\nu}}^b \eth {G^{\mu}}_b, \label{hg}
\ee
where ${G^\mu}_a
(x^a, \zeta, \zetabar)$ is the parallel propagator that takes vectors
from the point $x^a$ to $\scrip$ along the null geodesic $\ell_x(\zeta,
\zetabar)$, and ${A^{\mu}}_{\nu}$ are the asymptotic values of the
components of the Christoffel symbol in the direction of the connecting
vector on $\scrip$.  See \cite{Lamberti} and \cite{svi1} for a
derivation of this result.

It will be convenient (see (59)) to express the above relation in terms of a
given {\it
null} tetrad ${\lambda^a}_i (x)$ defined on $\cal M$ and $\scrip$ with
normalization
\be
g_{ab} {\lambda^a}_{\ihat} {\lambda^b}_{\jhat} =
\eta_{{\ihat} {\jhat}},
\ee
with
\be
\eta_{{\ihat} {\jhat}} =
\eta^{{\ihat} {\jhat}} = \left( \begin{array}{cccc} 0 & 1 & 0 & 0 \\
1 & 0 & 0 & 0 \\
0 & 0 & 0 & -1 \\
0 & 0 & -1 & 0 \end{array} \right).
\label{mtrxeta}
\ee
and ${H^{\ihat}}_{\jhat} = {H^{\mu}}_{\nu} {\lambda^{\ihat}}_{\mu}
{\lambda^{\nu}}_{\jhat}$, etc.
Tetrad indices are raised and lowered with
$\eta^{{\ihat} {\jhat}}$ and $\eta_{{\ihat} {\jhat}}$, respectively.
For the tetrad fields ${\lambda^\mu}_{\ihat}$ at $\scrip$ we will use
the Bondi tetrad $ \{ \ell_\mu, n_\mu, m_\mu, \mbar_\mu \} $ associated
with the Bondi coordinates.

The Bianchi Identities (37) can be written in terms of the holonomy operator
as
\be
\ethbar (H-A)^{\ihat} {}_{\jhat} - \eth
{(\Hbar-\Abar)}^{\ihat} {}_{\jhat} + [ H - A,  \Hbar - \Abar ]
{}^{\ihat} {}_{\jhat} - [ H, \Abar ] {}^{\ihat} {}_{\jhat} = 0.
\label{allbis}
\ee

A derivation of this result, via an integration of (37) over the region $V$ of
Figure 3, can be
found in \cite{thesis}.  Using $\eta^{{\ihat} {\jhat}}$ to raise one of
the indices, we obtain $H^{{\ihat} {\jhat}}$, $A ^{{\ihat} {\jhat}}$,
and their complex conjugates. Since these are skew in the ${\ihat}
{\jhat}$ indices, they can be separated into self-dual and
anti-self-dual parts, yielding
\be
H^{{\ihat} {\jhat}} =
H^{(-)}{}^{{\ihat} {\jhat}} + h^{(+)}{}^{{\ihat} {\jhat}} \; \; {\rm
\ and \ } \; \; A^{{\ihat} {\jhat}} = A^{(+)}{}^{{\ihat} {\jhat}} +
A^{(-)}{}^{{\ihat} {\jhat}}
\ee
and
\be
\Hbar^{{\ihat} {\jhat}} =
\Hbar^{(+)}{}^{{\ihat} {\jhat}} + \lhbar^{(-)}{}^{{\ihat} {\jhat}} {\rm \ and
\ } \Abar^{{\ihat} {\jhat}} = \Abar^{(-)}{}^{{\ihat} {\jhat}} +
\Abar^{(+)}{}^{{\ihat} {\jhat}}.
\ee
We have three complex non-trivial
components of $H^{(-)}$ which we denote by $H_\alpha$ and likewise three
$h_\alpha$, where $\alpha =\{1, 2, 3\}$, as shown in Table 1.  As for
the characteristic data, the $A$'s, most of them are zeroes and ones: they are
given by the
following:
\be
{A^{i}}_{j} = \left(
\begin{array}{cccc} 0 & 0 & -1 & 0 \\
0 & 0 & 1 & -\bondidot \\
-\bondidot & 0 & 0 & 0 \\
1 & -1 & 0 & 0 \end{array} \right),
\ee
and
\be {{\Abar}^{i}}{}_{j} = \left( \begin{array}{cccc} 0 & 0 & 0 & -1 \\
0 & 0 & -\bondidotbar & 1 \\
1 & -1 & 0 & 0 \\
 -\bondidotbar & 0 & 0 & 0 \end{array} \right).
\ee

The self-dual and anti-self-dual parts of $H$ can be written out
explicitly as well,
{\small \be H^{(-){\ihat} {\jhat}} =
\left(
\begin{array}{cccc} 0 & {1\over{2}}(H^{01} - H^{+-}) & H^{0+} & 0 \\
-{1\over{2}}(H^{01} - H^{+-}) & 0 & 0 & H^{1-} \\ -H^{0+} & 0 & 0
& -{1\over{2}}(H^{01} - H^{+-}) \\
0 & -H^{1-} & {1\over{2}}(H^{01} - H^{+-}) & 0 \end{array} \right)
\ee}
and
{\small \be h^{(+){\ihat}
{\jhat}} = \left( \begin{array}{cccc} 0 & {1\over{2}}(H^{01} + H^{+-})
& 0 & H^{0+} \\
-{1\over{2}}(H^{01} + H^{+-}) & 0 & H^{1+} & 0 \\
0 &
-H^{1+} & 0 & {1\over{2}}(H^{01} + H^{+-}) \\ -H^{0-} & 0 &
-{1\over{2}}(H^{01} + H^{+-}) & 0 \end{array} \right),
\ee}
and likewise for $\Hbar$.

In order to raise and lower the ${\ihat},  {\jhat}$ indices, we have to
use the null-coordinate version of the Minkowski metric $\eta_{{\ihat}
{\jhat}}$ and $\eta^{{\ihat} {\jhat}}$.  For example,
\bea
{H^0}_1 &=&
H^{00} = 0, {\rm \ by \ skew \ symmetry \ }, \\ {H^0}_{+} &=& -
H^{0-}.
\eea
Also note that the complex conjugate of a quantity, say,
$\overline{H^{0-}}$ gives us $\Hbar^{0+}$: the $0$ and $1$ indices are
insensitive to complex conjugation, while $+$ and $-$ are
interchanged.  The following table displays our notation.

\vspace{.5in} \begin{center}

\begin{tabular}{||c|c|c||c|c|c||} \hline {new } & {$\ihat \jhat$} &
{spin} & {new} & {$\ihat \jhat$} & {spin} \\ {notation} & {component} &
{weight} & {notation} & {component} & {weight} \\ \hline $H_1$ & $-
H^{(-)} {}^{0+}$ & $2$ & $\Hbar_1$ & $- H^{(+)} {}^{0-}$ & $-2$
\\ \hline $H_2$ & $- H^{(-)} {}^{01}$ & $1$ & $\Hbar_2$ & $- H^{(+)}
{}^{01}$ & $-1$ \\ \hline $H_3$ & $- H^{(-)} {}^{1-}$ & $0$ & $\Hbar_3$
& $- H^{(+)} {}^{1+}$ & $0$ \\ \hline $h_1$ & $- h^{(-)} {}^{0+}$ & $0$
& $\lhbar_1$ & $- h^{(+)} {}^{0-}$ & $0$ \\ \hline $h_2$ & $- h^{(-)}
{}^{01}$ & $-1$ & $\lhbar_2$ & $- h^{(+)} {}^{01}$ & $1$ \\ \hline
$h_3$ & $- h^{(-)} {}^{1-}$ & $-2$ & $\lhbar_3$ & $- h^{(+)} {}^{1+}$ &
$2$ \\ \hline \end{tabular}

\end{center}

\vspace{0.1in} \begin{center} Table 1: Notation.  \end{center}
\vspace{.5in}

With this notation the holonomy Bianchi identities (\ref{allbis})
become:
\bea \eth h_1-\ethbar H_1 + 2 h_1 H_1 -2 H_1 h_2 + 2 H_2 & =
& 0, \label{bi1}\\ \eth h_2 - \ethbar H_2 +h_3 H_1 -H_3 (h_1+1)+h_1 &
= & - \bondidotbar H_1, \label{bi2} \\ \eth h_3 - \ethbar H_3 + 2 h_2
(H_3-1)-2 H_2 h_3 &=& -\eth  \bondidotbar +2  \bondidotbar H_2,
\label{bi3}
\eea
where we have the data or ``driving'' terms (i.e.,
those that involve $\bondibar$) on the right.  Equation (\ref{allbis})
actually contains this triplet as well as their conjugates. They are the
result of taking the self-dual and anti-self-dual parts of equation
(\ref{allbis}).  This is the first set of equations that was given
symbolically as equation (\ref{symbbi})  in Section I. (\.  Note that
the first equation is algebraic in $h_2$ and $H_2$, and the second is
algebraic in $h_3$ and $H_3$.  We will exploit this structure in
Section III.D where we study these equations in more detail.



\subsection{The $\Lambda(H)$-relations.}

In the previous Section we defined the parallel propagator, ${G^\mu}_a
(x^a, \zeta, \zetabar)$, which takes vectors from a point on $\scrip$
to an interior point $x^a$ along a null geodesic $\ell_x(\zeta,
\zetabar)$.  In particular, the tetrad fields $\{
{\lambda^{\ihat}}_{\mu}, {\lambda^\mu}_{\ihat} \}$ that are defined on
$\scrip$ can be parallelly propagated in this manner to the point
$x^a$.  The parallelly propagated tetrad fields at $x^a$ are given by
\be {e^{\ihat}}_a = {\lambda^{\ihat}}_{\mu} {G^\mu}_a, \ee where
${\lambda^{\ihat}}_{\mu}$ represents the tetrad $ \{ \ell_\mu, n_\mu,
m_\mu, \mbar_\mu \}$ on $\scrip$.  We use the following notation for
the parallelly propagated tetrad at $x^a$:  \be {e^{\ihat}}_a \equiv \{
{e^0}_a, {e^1}_a, {e^{+}}_a, {e^{-}}_a \} \equiv \{ \ell_a, n_a, m_a,
\mbar_a \}.  \ee

If we take two sets of tetrads on $\scrip$, one at $(\zeta, \zetabar)$
and the other (on the same cut, associated with $x^a$) at $(\zeta +d
\zeta, \zetabar)$, and parallelly propagate both to the point $x^a$,
the difference in terms of the holonomy operator $H$ and the free data
$A$ \cite{Lamberti} is given by \be {e^a}_{\jhat} \eth {e^{\ihat}}_a =
{H^{\ihat}}_{\jhat} - {A^{\ihat}}_{\jhat} \> \>, \label{hhat} \ee \be
{e^a}_{\jhat} \ethbar {e^{\ihat}}_a = \Hbar {}^{\ihat} {}_{\jhat} -
\Abar {}^{\ihat} {}_{\jhat}.  \label{hbarhat} \ee

We have introduced three different tetrad bases:
${\lambda^{\ihat}}_a$, ${\lambda^{\ihat}}_{\mu}$, and
${e^{\ihat}}_a$.   The first of these, ${\lambda^{\ihat}}_a$, is an
arbitrary tetrad given at $x^a$ and therefore independent of $\zeta$
and $\zetabar$. The second, ${\lambda^{\ihat}}_{\mu}$, is a $(\zeta,
\zetabar)$-dependent Bondi tetrad given on $\scrip$. And the third,
${e^{\ihat}}_a$, is the result of parallelly propagating the
${\lambda^{\ihat}}_{\mu}$ in from $\scrip$ to $x^a$ along
$\ell_x(\zeta, \zetabar)$, and therefore is also $(\zeta,
\zetabar)$-dependent.

{}From Section III, we have at our disposal the $\theta^{i}$ coordinates
and the associated bases
${\theta^{i}}_a$ and ${\theta^a}_{i}$ at $x^a$. Since any vector
at $x^a$ can be expressed as a linear combination of either of the two
sets $ \{ {\theta^{i}}_a \}$ and $ \{ {e^{\ihat}}_a \}$, we can go from
one to the other, using the invertible transformation
\be
{\theta^i}_a =
{\Sigma^i}_{\ihat} {e^{\ihat}}_a. \label{bigsig}
\ee

We choose one of the `legs' of the
parallelly propagated tetrad vectors, ${e^0}_a$, equal to the vector
$\nabla_a Z$, i.e.,
\be
{e^0}_a \equiv \ell_a = {Z,}_a \equiv
{\theta^0}_a. \label{e0theta0}
\ee
Using this assumption and equations
(\ref{hhat}) and (\ref{hbarhat}) with our notation from Table 1, the
transformation matrix ${\Sigma^i}_{\ihat} \equiv {\bf \Sigma}$ and its
inverse ${\Sigma^{\ihat}}_i \equiv {\bf \Sigma^{-1}}$ can be
calculated (by application of $\eth$ and $\ethbar$ to (59) and (60)) and
written explicitly as functions of the $H$'s and $h$'s
\cite{thesis}.  $\bf \Sigma$ is given by
\be
\left(
{{\Sigma}^i}_{\ihat} \right) = \left( \begin{array}{cccc} 1 & 0 & 0 & 0 \\
{{\Sigma}^{1}}_0 & {{\Sigma}^{1}}_1 & {{\Sigma}^{1}}_{+} &
{{\Sigma}^{1}}_{-}\\
-(H_2+\lhbar_2) & 0 & (1+\lhbar_1) & H_1\\
-(\Hbar_2+h_2) & 0 &  \Hbar_1 & (1+h_1) \end{array} \right)
\ee with
{\small
\bea {{\Sigma}^{1}}_0 &=& -\eth (\Hbar_2+h_2) +
(\Hbar_2+h_2)(H_2+\lhbar_2) +  \Hbar_1 (\bondidot + \lhbar_3)+
(1+h_1)(H_3-1) \nonumber \\ {{\Sigma}^{1}}_1 &=& (1+h_1) (1+\lhbar_1) +
H_1 \Hbar_1 \nonumber \\ {{\Sigma}^{1}}_{+} &=& \eth \Hbar_1 -
(\Hbar_2+h_2) (1+\lhbar_1) - \Hbar_1 (H_2-\lhbar_2) \nonumber \\
{{\Sigma}^{1}}_{-} &=& \eth \lhbar_1 + (H_2-\lhbar_2) (1+h_1) - H_1
(\Hbar_2+h_2) \nonumber \eea}

Using $\bf \Sigma$, its conjugate, the equations (\ref{etheta}),
i.e.,
\be
\eth {\theta^{i}}_a = {T^{i}}{}_{j} {\theta^{j}}_a
\ee
and
their conjugates, the  ${T^{i}}{}_{j}$'s and ${\Tbar^{i}}{}_{j}$'s of Section
II can be
expressed directly in terms of the $H$'s and $h$'s via the ${\Sigma^{i}}{}_{j}$
by

\be
\bf T = (\eth \Sigma) \cdot {\Sigma}^{-1} + \Sigma
\cdot ( H-A ) \cdot {\Sigma}^{-1},\label{Th}
\ee
where the boldface represents the corresponding matrices \cite{thesis}. Since
${T^{i}}{}_{j}$ was shown in equation
(\ref{Tmatrix}) to have the form
\be
{\bf T} \equiv ({T^{i}}_{j}) =
\left( \begin{array}{cccc} 0 & 0 & 1 & 0 \\ {T^1}_0 & {T^1}_1 &
{T^1}_{+} & {T^1}_{-} \\ \Lambda_0 & \lone & \Lambda_{+} & \Lambda_{-}
\\ 0 & 1 & 0 & 0 \end{array} \right).
\ee
we obtain from (\ref{Th}) and (66) the
$\Lambda(H)$-relations:
\bea \Lambda_1 &\equiv& \lone = {{2(\lhbar_1+1) H_1} \over
{(\lhbar_1+1)(h_1+1)+ \Hbar_1 H_1}}, \label{lambda1}\\ \Lambda_+ &=& W
- {1 \over{2}} \left( \Lambda_1 \Wbar + \ethbar \Lambda_1 - \eth \ln
q \right), \label{lambdap} \\ \Lambda_- &=& {1 \over{2}} \left[ \eth
\Lambda_1 - \Lambda_1 {{\eth(H_1 \Hbar_1)} \over{H_1 \Hbar_1}} -
\Lambda_1 \eth \ln \left( {q \over{1-\sqrt{q}}} \right)
\right],\label{lambdam} \\ \Lambda_0 &=& \bondidot + {1 \over{2}}
\Lambda_1 - {1 \over{4}} \eth \ethbar \Lambda_1 + O(H^2),
\label{lambda0} \eea where $W$ is defined by \be W \equiv {{(\eth
\lone - 2 \Lambda_{-})} \over{\lone}} - \eth \ln q, \nonumber \ee with
\be q \equiv (1 - \lone \lonebar). \nonumber \ee See \cite{thesis} and
\cite{YM} for a more complete derivation of these equations. The
precise form of equation (\ref{lambda0}) turns out to be rather lengthy
and is given in Appendix A of \cite{thesis}.


\subsection{The Holonomy Field Equations.}

In this section, we describe an important result obtained earlier
\cite{Lamberti}, wherein the vacuum Einstein equations in the form \be
^{+}\!F_{ab}^- = 0 \ee were imposed on the holonomy operator.  The goal
is to obtain field equations for the holonomy operator, and therefore
we first need a relationship between the curvature tensor and the
holonomy operator.

It can been shown \cite{Lamberti}, using a
non-Abelian version of Stokes' theorem, that the holonomy operator satisfies
\be
H = \int_{s_{\scriptscriptstyle 0}}^{\infty}(F_{ab}^+
+ F_{ab}^-) \;\ell^a M^b ds = h^{(+)} + H^{(-)} \label{hFab}
\ee
and
\be
\Hbar = \int_{s_{\scriptscriptstyle 0}}^{\infty}(F_{ab}^+ +
F_{ab}^-) \;\ell^a \Mbar^b ds = H^{(+)} + h^{(-)}, \label{hbarFab}
\ee
where the $H$ and $h$ with the plus and minus signs were defined in
Section III.B, $s$ is an affine parameter along the generators of the
light cone at $x^a$ with $s=s_{\scriptscriptstyle 0}$ at $x^a$.  $M^a$
and $\Mbar^a$ are connecting vectors between, respectively the long
legs of $\Delta_x$ and $\Deltabar_x$.  See \cite{Lamberti} for more
details.

The two equations (\ref{hFab}) and (\ref{hbarFab}) can be inverted to
obtain $F_{ab}^-$ and $F_{ab}^+$ explicitly in terms of the components
of the holonomy operator.  Specifically $F_{ab}^-$ and $F_{ab}^+$ can
be expressed as two derivatives of the $H$'s with respect to $\theta^1
= R$. Using this expression for the curvature in
$^{+}\!F_{ab}^- = 0$ one obtains a differential relationship between
the different components of $H$ and $\Hbar$, given by \be [q^{-1}
h_{\alpha,R}]_{,R} + \delta [q^{-1} \lone h_{\alpha,R}]_{,R} = [q^{-1}
\lonebar H_{\alpha,R}]_{,R} + \delta [q^{-1}H_{\alpha,R}]_{,R},
\label{rawEeqns} \ee where \be \delta = {{\sqrt{q} - 1} \over{\lone}}.
\ee

Note that the equations (\ref{rawEeqns}),{\it  which we refer to as the
field equations}, are three linear relations between the $H$ and $h$
with coefficients depending $Z$ via $\Lambda_1$.  They are our final
set of equations.  In the next section, we will use these with the
Bianchi identities and the $\Lambda(H)$-relations to simplify the
overall structure of the theory.


\subsection{The Full Theory.}

We have at this point three sets of equations:  \begin{itemize} \item
the three holonomy Bianchi identities, \bea \eth h_1-\ethbar H_1 + 2
h_1 H_1 -2 H_1 h_2 + 2 H_2 &=& 0, \label{bi1too}\\ \eth h_2 - \ethbar
H_2 +h_3 H_1 -H_3 (h_1+1)+h_1 &=& - \bondidotbar H_1, \label{bi2too} \\
\eth h_3 - \ethbar H_3 + 2 h_2 (H_3-1)-2 H_2 h_3 &=& -\eth
\bondidotbar +2  \bondidotbar H_2, \label{bi3too} \eea \item the
holonomy field equations, \be [q^{-1}h_{\alpha,R}]_{,R} + \delta
[q^{-1}\lone h_{\alpha,R}]_{,R} = [q^{-
1}\Lambdabar_{,R}H_{\alpha,R}]_{,R} + \delta [q^{-1}H_{\alpha,R}]_{,R},
\label{rawEeqns2} \ee \item the $\Lambda(H)$-relations,, \bea \Lambda_1
&\equiv& \lone = {{2(\lhbar_1+1) H_1} \over {(\lhbar_1+1)(h_1+1)+
\Hbar_1 H_1}}, \label{lambda1too}\\ \Lambda_+ &=& W - {1 \over{2}}
\left( \Lambda_1 \Wbar + \ethbar \Lambda_1 - \eth \ln q \right),
\label{lambdaptoo} \\ \Lambda_- &=& {1 \over{2}} \left[ \eth \Lambda_1
- \Lambda_1 {{\eth(H_1 \Hbar_1)} \over{H_1 \Hbar_1}} - \Lambda_1 \eth
\ln \left( {q \over{1-\sqrt{q}}} \right) \right], \label{lambdamtoo} \\
\Lambda_0 &=& \bondidot + {1 \over{2}} \Lambda_1 - {1 \over{4}} \eth
\ethbar \Lambda_1 + O(H^2), \label{lambda0too} \eea \end{itemize} and
their conjugates.  The first set involves the $H$'s and $h$'s and the
data only, while the other two involve the $H$'s and $h$'s and the
$\Lambda_i$.  We can think of the $H$'s and $h$'s as describing the
``field" and the $\Lambda_i$, the ``background".  In this Section we
study the above equations and show how they can be reduced to a smaller
and simpler set.

This simplification procedure begins with the following important
observation: equation (\ref{lambda1too})  its complex conjugate are two
{\it algebraic} relations between the six quantities, viz.,  $\lone$
and $\lonebar$, $H_1$, $\Hbar_1$, $h_1$ and $\lhbar_1$.  Solving for
$h_1$ and $\lhbar_1$ algebraically yields,
\bea h_1 &=& -{H_1 \over
{\deltabar}}  -1,  \label{lhbigh1} \\
\lhbar_1 &=& -{\Hbar_1 \over
{\delta}}  -1, \label{lhbigh1bar}
\eea where
\be
\delta = {{\sqrt{q} - 1} \over{\lone}} \; \; {\rm and} \;\; q=1-\lone
\lonebar,  \label{delta}
\ee
Thus $h_1$ and $\lhbar_1$ are not independent quantities
and equations (\ref{lhbigh1}) and (\ref{lhbigh1bar}) allow us to
eliminate them completely.  This {\it algebraic} structure extends to
the other $h$'s.  We are able to solve, from equations
(\ref{bi1too}) and (\ref{bi2too}, for all the remaining $h$'s and completely
eliminate them from the remainder of the analysis.  The algebraic
relationships between $H_\alpha$ and $h_\alpha$ for all
$\alpha=\{1,2,3\}$ takes the form
\be
h_\alpha = -{H_\alpha \over
{\deltabar}} + G_\alpha,\label{lhbighalpha}
\ee
where $G_\alpha$ is a function of $\Lambda_1$ and the preceding $H$'s. See
\cite{thesis} for explicit expressions for the $G_\alpha$.  The
important consequence of the above structure is that after eliminating
the $h$'s from the holonomy field equations (\ref{rawEeqns2}) we obtain a very
attractive set of three differential equations for $H_\alpha$:
\be
{{\partial {}^2 H_\alpha} \over {\partial R^2}} + X {{\partial
H_\alpha} \over {\partial R}} + Y H_\alpha = \mu {{\partial {}^2
G_\alpha} \over {\partial R^2}} + \nu {{\partial G_\alpha} \over
{\partial R}},\label{Eeqns}
\ee
where $X$, $Y$, $\mu$ and $\nu$ are
functions only of $\Lambda_1$ and $\Lambdabar_1$.

These three equations are linear with the only difference among them
being in the homogeneous terms: the first of them, the equation for
$H_1$, is homogeneous with the coefficients depending only on
$\Lambda_1$ (a function of $R$), while for the next pair the
inhomogeneous terms are driven by the solutions of the previous ones.
{}From this structure, if the homogeneous equation could be solved, the
remaining ones could be solved by quadratures.

To summarize, we have so far used the information contained in two (of the
three) Bianchi identities, one of the $(\Lambda, H)$-equation, and all
the field equations.  The third Bianchi identity has not yet played any
role.  {\it Assuming} that we could solve (90) for $H_1$
as a function (functional) of $\Lambda_1$ we would have all the
$H_\alpha = H_\alpha(\Lambda_1, \Lambdabar_1)$ and their complex
conjugates.  These expressions, which would be explicit functions of
$\Lambda_1$ and $\Lambdabar_1$, could then be substituted in the third
Bianchi identity.   The resulting equation would involve only $\Lambda$,
 $\Lambdabar$, and the data.  The single resulting equation, involving only
$\Lambda_1$ and $\Lambdabar_1$, i.e.,
\bea
\eth (H_3/\deltabar + G_3)
- \ethbar H_3 + 2 (H_2/\deltabar + G_2) (H_3-1)-2 H_2 (H_3/\deltabar +
G_3) \nonumber \\ \;\;\;\;\;\;\;\;\; = -\eth  \bondidotbar +2
\bondidotbar H_2, \label{pide}
\eea
would be our sought for equation
for the determination of the characteristic surfaces and therefore the
conformal metric of the Einstein spacetime.

The equation, however, can be manipulated, from the leading terms (but not
explicitly) of the
remaining $\Lambda(H)$-relations, into a simpler and more attractive
equation that is referred to as the light cone cut equation (LCCE)
whose solution directly yields the light cone cut function.

Using the solution to equation (\ref{Eeqns}) in the form
\be
H_1 = {1
\over{2}} \Lambda_1 + O (\Lambda ^2) , \label{H1}
\ee
we have from substitution into equation (\ref{pide})
\be
\ethbar {}^3 \Lambda_1 = -4 \eth \bondidotbar + {\rm
\ higher \ order \ terms, \ } \label{ethbar3l1}
\ee
where $\bondi$ is
free data given on $\scrip$.  Manipulating this and using the equation
(\ref{lambda0too}) of the $\Lambda(H)$-relations we obtain,
\be
\ethbar {}^2 \Lambda_0 = \eth {}^2 \bondidotbar + \ethbar {}^2
\bondidot + {\rm \ higher \ order \ terms, \ }
\ee
or, from the
definition $\eth {}^2 Z_{,0} = \Lambda_0$,
\be
\ethbar {}^2 \eth
{}^2 Z = \eth {}^2 \bondibar + \ethbar {}^2 \bondi + {\rm \ higher
\ order \ terms \ }\label{2laplace}
\ee
as our final equation for the light cone cut function $Z$.  Although the higher
order terms have not been worked out explicitly, they can be obtained via a
perturbation scheme (see next Section). The solutions to equation
(\ref{2laplace}) yield the conformal structure of asymptotically flat
vacuum spacetime.

In the above analysis it was assumed that the solution to the equation
for $H_1(\Lambda)$, namely,
\be {{\partial {}^2 H_1} \over {\partial R^2}} + X
{{\partial H_1} \over {\partial R}} + Y H_1 = 0, \label{lode}
\ee
was
known.  Since it is a linear second order o.d.e. solutions for arbitrary
$\Lambda(R)$ must exist, though it is not clear--nor is it likely--that
we will be able to solve it explicitly. It will however, always be of
the form
\be H_1 = {1 \over{2}} \Lambda_1 + O (\Lambda ^2)
\ee
with the higher order terms computable perturbatively.

Since the LCCE, (\ref{2laplace}),
determines the conformal structure of vacuum spacetimes, the only
remaining quantity to be determined is the conformal factor needed to
convert the conformal metric into a vacuum metric.  In the $\theta^i$
coordinate system the relation between the conformal metric $ g$
and the full metric ${\hat g}$ is given by \be {\hat g}^{ij} = \Omega^2 {\it
g}^{ij}
\ee It can be shown \cite{thesis} that the conformal factor can be
expressed as a simple function of $H_1$ and $\Hbar_1$, namely, \be
\Omega^2 = (1+h_1) (1+\lhbar_1) + H_1 \Hbar_1.\label{confac} \ee
[As an
aside we remark that from equation (\ref{lode}), it is possible to show
that the conformal factor (\ref{confac}) satisfies the equation
\[
{d^2 \Omega \over{d R^2}} = Q(\Lambda) \Omega,
\]
where $Q = Q(\Lambda)$ is a simple known function of $\Lambda_1$. The above
equation is known in the literature \cite{oregon} as the Einstein bundle
equation.]


\section{An Iterative Scheme.}

Much of the understanding  of the structure and meaning of our final
equations has come from the application of an iterative scheme to the
formal theory and looking at the leading behavior.  In this section,
we present the results obtained in linear order.

We expand all quantities in powers of a small parameter $\epsilon$
which measures the deviation from flatness. From the assumption that
$\epsilon$ enters as a multiplicative factor of the Bondi shear, i.e.
via $\epsilon \sigma_B$, it becomes clear that the expansions have the
form \bea Z &=& \Zzero + \epsilon \> \Zone + \epsilon^2 \> \Ztwo  +...
,\\ e^i_a  &=&  {\ezero}^i_a +  \epsilon \> {\eone}^i_a +  \epsilon^2
\> {\etwo}^i_a +... ,\\ \Lambda &=& \epsilon \> {\Lambdaone} +
\epsilon^2 \> {\Lambdatwo} + \epsilon^3 \> {\Lambdathree}... , \\ H
&=&  \epsilon \> {\Hone} +  \epsilon^2 \> {\Htwo} + \epsilon^3 \>
{\Hthree}... ,\\ h &=&  \epsilon^2 \> {\htwo} +  \epsilon^3 \>
{\hthree} + ... .  \eea There is no zeroth order contribution to
$\Lambda$ or the holonomy operator since these quantities are zero for
Minkowski space.  Furthermore,  a direct calculation (using equations
(\ref{lhbigh1}),  (\ref{delta}), and (\ref{H1})) shows that $h$ begins at
second order.

Since $\Lambda$ starts with order one in the perturbation expansion, it follows
from the relationship between $Z$ and $\Lambda$ that $\Zzero$ satisfies
\be \eth{}^2 {\Zzero}_a = 0,
\ee
whose solution is the Minkowski space light cone cut
function
\[ \Zzero = x^a {\ell\!\!\!^{^{\scriptscriptstyle {(0)}}}}_a,
\] with
\[ {\ell\!\!\!^{^{\scriptscriptstyle {(0)}}}}_a
(\zeta,\zetabar) = \frac{1}{\sqrt{2} P} (1+\zeta\zetabar,
\zeta+\zetabar,i(\zeta-\zetabar), -1+\zeta\zetabar); \;\; \; P
=(1+\zeta\zetabar).  \]


\subsection{Linearized Gravity.}

We begin the linearization with the three holonomy Bianchi identities,
which to first order are:  \bea {\Hone}_2 &=& {1 \over{2}} \ethbar
{\Hone}_1, \label{bi1one}\\ {\Hone}_3 &=& - \ethbar {\Hone}_2,
\label{bi2one}\\ \ethbar {\Hone}_3 &=& \eth
\bondidotbar\label{bi3one}, \eea where the $h$'s have disappeared since
they have no first order contribution.  The next step in the process of
elimination would naturally be to solve the holonomy field equations
obtaining the $H$'s in terms of the $\Lambda$.  However, to first (and
second order) we can circumvent this step and consider the first
of the $(\Lambda, H)$-equations directly.  It leads to an expression
for $H_1$, which with equations (\ref{bi1one}) and (\ref{bi2one}),
yields the first order $H_{\alpha}$ that automatically satisfy the
field equations (\ref{Eeqns}) to first order.  (For third and higher
order we would have to solve the field equations {\it before} using the
$\Lambda(H)$-relations for the simplification procedure.)

We thus have \be {\Hone}_1 = {1 \over{2}} {\Lambdaone}_1\label{H1one},
\ee \be {\Hone}_2 = {1 \over{4}} \ethbar {\Lambdaone}_1, \label{H2one}
\ee \be {\Hone}_3 = - {1 \over{4}} \ethbar {}^2 {\Lambdaone}_1.
\label{H3one} \ee Our final equation for the cut function is the
linearized third Bianchi identity, which, after substituting for $H_3$,
is \be \ethbar {}^3 \Lambda_1 = -4 \eth \bondidotbar.
\label{thirdbione} \ee (Although we did not have to solve the holonomy
field equations, it is a straightforward but slightly tedious process to
verify that the expressions (\ref{H1one})-(\ref{H3one}) for the $H$'s
do satisfy the field equations.  A minor subtlety in the calculation is
that the linearization of the field equations yields identically
vanishing expressions, so that one must look at the equations at their
first {\it non-vanishing} order.  Performing the calculation,
nevertheless, serves as a consistency check between the different sets
of equations.)

At this point we have used all the equations except the three remaining
$\Lambda(H)$-relations, \bea \Lambda_{-} &=& {1 \over{2}} \eth
\Lambda_1, \label{lmone}\\ \Lambda_{+} &=& - {1 \over{2}} \ethbar
\Lambda_1, \label{lpone}\\ \Lambda_0 &=& \bondidot + {1 \over{2}}
\Lambda_1 - {1 \over{4}} \eth \ethbar \Lambda_1.  \label{lzone} \eea
These relations can be used to write (\ref{thirdbione}) in the more
symmetric form as the linearized version of the light cone cut
equation:  \be \ethbar{}^2 \eth{}^2  Z  = \eth{}^2 \bar{\sigma}_B(Z)
+ \ethbar{}^2 \sigma_B(Z),\label{Mone} \ee which is the light cone cut
equation accurate to first order.  In other words, this equation is
equivalent to the the linearized conformal Einstein equations.  (The
above equation for linear theory has been derived from the vanishing of
the Bach tensor, by Lionel Mason\cite{mason}.)

In linear theory (see(99))the conformal factor equals 1, and therefore in this
approximation we have not only the conformal Einstein equations but the
Einstein equations themselves.

Equation (\ref{Mone}), is to be thought of as an equation for $Z$,
whose solution can be written as a sphere integral using the Green
function G of the operator $\ethbar{}^2 \eth{}^2 $.  See Appendix C of
\cite{thesis} for a derivation of the Green function of this operator.
In other words we can write
\be Z(x,\zeta,\zetabar)=\int{ G(\zeta, \zetabar; \eta, \etabar)
[\eth{}^2 \bondibar(\Zzero(x,\eta,\etabar)) + \ethbar{}^2
\bondi(\Zzero(x,\eta,\etabar)) ] dS_{\eta}},
\ee
with $dS_{\eta}$ the sphere volume element in $(\eta,\etabar)$
coordinates, as the general solution to the asymptotically flat
linearized vacuum Einstein equations.  In principle, though we have not
yet done so in practice, the higher order terms could be calculated
successively by similar integrals but with the integrands depending on
terms to lower order in the perturbation scheme.


\section{Summary and Conclusions.}

We have obtained three sets of coupled equations, the holonomy Bianchi
Identities, the $\Lambda(H)$-relations and the ``field equations" for
the holonomy operator $H$ and the light cone cut function $Z$, which
are equivalent to the full vacuum Einstein equations.  These equations,
which already have built into them the free choice of Bondi data
$\bondi (u, \zeta, \zetabar)$, can be manipulated and simplified (in
structure) to one (complicated) equation, the LCCE, and a simple one
for the conformal factor.  On analysis, they yield perturbatively, a
D'Adhemar-like formulation of general relativity.

On the negative side these equations are quite unusual and are based
on unfamiliar ideas and variables, and unfortunately are quite complicated.
They
nevertheless have some features of considerable attractiveness: new
insights often can be gained from the use of new variables; the
perturbative solutions, from given data, are essentially unique; as the
perturbation calculation proceeds to higher order the formalism yields
the corrected light cone structure from the preceding order, in
contrast to the usual perturbation theory which uses, at all orders, the
Minkowski light cone structure.

As a final comment, we remark that the work reported on here is the
direct antecedent of another approach to general relativity (the Null
Surface Theory of GR, to be reported elsewhere) that is based solely on
families of characteristic surfaces as the basic variable of the
theory--without any mention of fields or holonomies.  The new view,
which has certain similarities to the present work and will probably
yield considerable simplifications over it, could not have been
developed without the current approach.

\vspace{1cm}

{\large {\bf Acknowledgements}}: This work was written under a collaboration
grant from the NSF and CONICET. CNK and ETN thank their support.
\vspace{1cm}

{\large {\bf References}}

\end{document}